# Ferroelectric and Photovoltaic Properties of Transition Metal doped Pb(Zr$_{0.14}$Ti$_{0.56}$Ni$_{0.30}$)O$_{3-\delta}$ Thin Films

Shalini Kumari[1], Nora Ortega[1], Ashok Kumar[2,#], J. F. Scott[1,3], R. S. Katiyar[1,#]

[1]Department of Physics and Institute for Functional Nanomaterials, University of Puerto Rico, San Juan, PR 00931-3334, USA
[2]CSIR-National Physical Laboratory, New Delhi-110012, India
[3]Department of Physics, Cavendish Laboratory, University of Cambridge, Cambridge CB3 OHE, United Kingdom

**Abstract:**

We report nearly single phase Pb(Zr$_{0.14}$Ti$_{0.56}$Ni$_{0.30}$)O$_{3-\delta}$ (PZTNi30) ferroelectric having large remanent polarization (15-30 μC/cm$^2$), 0.3-0.4 V open circuit voltage (V$_{OC}$), reduced band gap (direct 3.4 eV, and indirect 2.9 eV), large ON and OFF photo current ratio, and the fast decay time. Reasonably good photo current density (1-5 μA/cm$^2$) was obtained without gate bias voltage which significantly increased with large bias field. Ferroelectric polarization dictates the polarity of V$_{OC}$ and direction of short circuit current (I$_{SC}$), a step forward towards the realization of noncentrosymmetric ferroelectric material sensitive to visible light.

#Corresponding Authors: Ram S Katiyar (Email: rkatiyar@uprrp.edu), Ashok Kumar (ashok553@gmail.com)

**Introduction:**

Ferroelectric semiconductors and their bandgap engineering represent the most fascinating research area since the discovery of ferroelectricity and related phenomenon. [1,2] Recently it has been experimentally observed that solid-solution of potassium-niobate and barium-nickel-niobate perovskite ferroelectrics possess immense potential for photovoltaic (PV) applications with reduced band gap and moderate quantum efficiency [3]. Generally, ferroelectric materials are high bandgap insulators with low leakage current system. A long-standing challenge in solid state physics is the tailoring of bandgap of ferroelectric host matrix with transition metal ions at B-site, which in turn keeps the polarization intact with an enhancement of the bulk photovoltaic (PV) effect. [4,5,6,7] As we know in general, organic and other semiconconductor based PV cells require p-n junctions for the creation of photo-induced charge carriers, and hence the limitation of these devices is that it cannot produce open circuit voltage ($V_{OC}$) above the band gap of the materials. In these systems p-n junction is a property of the interface and not of a bulk property of materials. However, ferroelectric photovoltaic systems have unique natural properties, including granularity and non-centrosymmetry, and hence these do not need any p-n junction for photo currents. They can also produce exceptionally large $V_{OC}$ far above their bandgap for in-plane configuration with domains and domains walls manipulation.[8] Recently, Yang et al.[9] have shown the above band gap $V_{OC}$ by tailoring the in-plane domains and domain walls in $BiFeO_3$ (BFO) thin films, which is relatively small bandgap ($E_g \sim$ 2.6-2.9 eV) ferroelectric semiconductor. The basic mechanism of the domain-wall-based ferroelectric PV is quite different from that of inversion center-symmetry absence in the bulk ferroelectric. [8,10]

Bennett et al. [11,12] had utilized first principle density functional theory (DFT) calculations on the solid solution of PbTiO$_3$ (PTO) and Ba(Ti$_{1-x}$Ce$_x$)O$_3$ (BTCO) with partial substitution of different transition metal cations, and they predicted that removal of 50 % Ti ions or more can lower the band gap below 1 eV with remanent polarization comparable to that of pure PTO and BTCO. This particular prediction can lead the design and discovery of new low-bandgap semiconductor ferroelectrics. However, in reality it would be difficult to produce single phase complex systems of lead titanate with transition metal cations. Many other ferroelectric materials, such as Pb(Zr$_{1-x}$Ti$_x$)O$_3$ [13,14], LiNbO$_3$ [15] and BaTiO$_3$ [16] also exhibit photoelectric and photovoltaic effects under illumination of visible and near ultraviolet light; but the magnitude of photo current and voltage obtained for the device application are far below the photo-electronics requirements. In this respect ferroelectric BFO with its very high polarization ~ 90 μC/cm$^2$ [17] and a direct band gap ~2.67eV [18] had shown tremendous potential for such optoelectronic applications. [5,19,20]

Pintilie et al.[3] reported band gap in Pb(Zr$_{1-x}$Ti$_x$)O$_3$ system that increased with Zr content from 3.9 eV to 4.4 eV. A lower bandgap value (3.9 eV) and larger photocurrent signal was obtained for Pb(Zr$_{0.20}$Ti$_{0.80}$)O$_3$.[3,21] Theoretical predictions of low bandgap highly polar semiconducting Pb(Ti$_{1-x}$Ni$_x$)O$_{3-x}$ stimulated us to check the experimental performance of similar systems.[4] We report fabrication of polycrystalline and highly grain-oriented Pb(Zr$_{0.20}$Ti$_{0.80}$)$_{0.70}$Ni$_{0.30}$O$_{3-\delta}$ (or in a more compact formula Pb(Zr$_{0.14}$Ti$_{0.56}$Ni$_{0.30}$)O$_{3-\delta}$) films on indium tin oxide (ITO)/Glass and La$_{0.67}$Sr$_{0.33}$MnO$_3$)/LaAlO$_3$ (LSMO/LAO) substrates. Both systems were found to be polar with reduced bandgap and reasonably good PV effects. Photo current switching dynamics and transient current behavior are also discussed.

**Experimental Details:**

The ceramic target of $Pb(Zr_{0.14}Ti_{0.56}Ni_{0.30})O_{3-\delta}$ (PZTNi30) was prepared by a conventional solid state reaction route. Analytical-purity oxides of PbO, $ZrO_2$, $TiO_2$, and NiO (Alfa Aesar) with purity of 99.99% were used as raw materials. The powders of the respective metal oxides were mixed in a planetary high-energy ball mill with tungsten carbide media. The milled powder was then calcined at 1100 °C for 10 h in a closed alumina crucible; 10% excess of PbO was added to compensate Pb-deficiency during the high temperature processing. The calcined powder was pressed to one-inch pellets and sintered at 1150 °C for 4 h. All heat treatments were performed in air medium. Sintered targets were used for the fabrication of high-purity PZTNi30 thin films on conducting $La_{0.67}Sr_{0.33}MnO_3$ (LSMO)/$LaAlO_3$ (LAO) (100) and ITO/glass substrates by pulse laser deposition. The growth conditions were as follows: (i) the substrate was kept at 600 °C, (ii) oxygen pressure ~ 80 mTorr, (iii) laser energy 1.5-2.5 J/cm$^2$, (iv) excimer laser (KrF, 248 nm) and (v) pulse frequency 5 Hz. After deposition, the as-grown films were annealed in pure oxygen at 300 Torr for 30 minutes at 700 °C and then cooled down to room temperature slowly. Similar conditions were used to grow LSMO layer on the LAO substrate and PZTNi30 on the ITO/glass substrates.

The orientation and phase purity of these films were examined at room temperature by x-ray diffraction systems (Siemens D5000 and Rigaku Ultima III) using CuK$_\alpha$ radiation with wavelength of $\lambda = 1.5405$ Å. Room-temperature topography and domain images of these thin films were recorded by Piezo force microscopy (PFM) (Veeco) operated in contact mode and using an ultra-sharp silicon tip with a resonance frequency of about 25 kHz. The film thickness was determined using an X-P-200 profilometer and filmetrics. To investigate the electrical properties square capacitors were fabricated by dc sputtering with semi transparent Pt top electrodes with area of ~$10^{-4}$ cm$^2$ utilizing a shadow mask. Frequency dependence of the dielectric and ferroelectric properties were measured using an HP4294A impedance analyzer and Radiant tester respectively at room temperature. Photovoltaic current was measured using solar simulator and Keithley-2401 at room temperature.

**Results and Discussion:**

Fig.1 and Fig.1 (inset) show the room temperature X-ray diffractograms (XRD) of PZTNi30/LSMO/LAO(100) and PZTNi30/ITO/glass heterostructure over 20-60 degree of Braggs angle. The θ-2θ scan of PZTNi30/LSMO/LAO(100) shows highly oriented film along (100) plane with small peaks along (101) (~ 8%) and (111) (~ 2%) directions for LAO substrate, whereas figure 1 (inset) illustrates polycrystalline nature of films grown on ITO coated glass. The XRD data were used to evaluate the lattice parameters of polycrystalline and oriented films of PZTNi30 on the basis of a tetragonal unit-cell of Pb(Zr$_{0.20}$Ti$_{0.80}$)O$_3$.[22,23] For comparison, bulk lattice parameters of PZTNi30 were also calculated from the ceramic pellet. All calculations were carried out using the UnitCellWin program. The results are listed in Table 1. In both cases, PZTNi30 films have a tetragonal crystal structure with a reduction of the tetragonality ($c/a$), i.e. c/a for polycrystalline and oriented films was ~1.003, comparatively smaller than the host matrix ($c/a = 1.041$).

Surface topography and domains switching of the films were investigated by the conducting mode atomic force microscopy (AFM) and piezo force microscopy (PFM) respectively. AFM images revealed that average size of grains for PZTNi30/LSMO/LAO(100) heterostructures are less than 500 nm with average surface roughness ~ 5.5 nm (see Fig. 2(a)) however, bigger grains and high average surface roughness were observed for PZTNi30/ITO/glass structures (not shown). The surface morphology of the films displayed the negligible evidence of cracks, voids and defects over large area. PFM images disclose the domain configurations and domain imaging, domain writing, domain dynamics and evolution. Fig. 2(b) and 2(c) show PFM phase and amplitude images of PZTNi30/LSMO/LAO(100) under bias field with an area of 5 x 5 μm$^2$. Most of the ferroelectric domains are switched under the external bias electric field by PFM tips; however, the contrast of the PFM phase image under ± 9 V bias field suggests some of the domains remain in the different switching direction; this may be due to domain growth in different directions and is well supported by XRD patterns. The PFM phase and amplitude images confirm the switching of domains at nanoscale. Note that domains and domain walls are not defined and aligned in any particular directions such as 90°, 180°, or (109°/71° in BiFeO$_3$) [24] that make it difficult to do in-plain measurements.

Fig. 3 (a) and 3(b) show the bulk polarization and dielectric tunability (insets in Fig. 3) of both systems. A well saturated polarization hysteresis was obtained for both polycrystalline PZTNi30 on glass and the highly oriented film. The values of remnant polarization was ~ 15-30 μC/cm$^2$ which is comparatively matching with the other reports on the Pb(Zr$_{1-x}$Ti$_x$)O$_3$ system.[12] It indicates that the hypothesis of "retaining of polarization while comprehensive decrease in bandgap" seems more realistic in case of transition metal doped PZT. Both these systems show dielectric tunability with well shaped butterfly loops under the application of external electric field prove its polar/ferroelectric nature. : Polycrystalline films show higher dielectric constant compare to highly oriented PZTNi30/LSMO/LAO(100) films. A small difference in the values of coercive field in oriented films was observed with two different probe techniques such as capacitance-voltage and polarization-voltage. This may be due to detection limits of displacement current and leakage current by two different apparatus; however, further studies needed to clarify this minor discrepancy.

Direct and indirect bandgaps of PZTNi30 were determined from the UV-visible transmission data. The direct band gap, $E_g$, was estimated from the modified square law using $(\alpha h\nu)^2$ versus $h\nu$ plots derived from the Tauc's relation[25,26],

$$(\alpha h\upsilon)^2 = A_{Tauc}(h\upsilon - E_g) \ldots\ldots\ldots\ldots\ldots\ldots\ldots\ldots\ldots\ldots\ldots\ldots\ldots\ldots\ldots\ldots\ldots\ldots\ldots\ldots\ldots\ldots\ldots\ldots\ldots\ldots(1)$$

Where, absorption coefficient $\alpha$ is defined:

$$\alpha = \frac{1}{d}\ln\left(\frac{100}{\%T}\right)\ldots\ldots\ldots\ldots\ldots\ldots\ldots\ldots\ldots\ldots\ldots\ldots\ldots\ldots\ldots\ldots\ldots\ldots\ldots\ldots\ldots\ldots\ldots\ldots\ldots(2)$$

Where, $d$ is the film thickness, $\%T$ is the percentage of transmission, $h\nu$ photon energy, and $E_g$ band gap. $A_{Tauc}$ (Tauc parameter) is the slope of the linear region in a plot of $(\alpha h\nu)^2$ vs. $h\upsilon$, whose extrapolation to $(\alpha h\upsilon)^2 = 0$ would give the value of the direct bandgap.

On the other hand, data from indirect bandgaps meet usually the Tauc's law:

$$(\alpha h\upsilon)^{1/2} = B_{Tauc}(h\upsilon - E_g) \ldots\ldots\ldots\ldots\ldots\ldots\ldots\ldots\ldots\ldots\ldots\ldots\ldots\ldots\ldots\ldots\ldots\ldots\ldots\ldots\ldots(3)$$

Where, $B_{Tauc}$ (Tauc parameter) is the slope of the linear region in a plot of $(\alpha h\nu)^{1/2}$ vs. $h\upsilon$, whose extrapolation to $(\alpha h\upsilon)^{1/2} = 0$ would give the value of the indirect band gap. It should be noted that these relationships are valid only for parabolic bands.

Fig. 4(a) presents the optical transmittance data for ITO/glass and PZTNi30/ITO/glass. The PZTNi30 layered structure on glass exhibits 72% transmittance at 600 nm, with a reduction of only 8% compared with pure ITO/glass substrate. This property is important since in photovoltaic applications transparent PZT is needed.[27] Fig. 4(b) and 4(c) show the Tauc's relation for the PZTNi30 thin films for both direct and indirect bands respectively; for comparison data from ITO/glass substrate is included in Fig 4(b). The direct band gap was calculated around 3.4 eV (less than 3.9 eV of pure $PbZr_{0.2}Ti_{0.8}O_3$),[3] indicating Ni-substitution at B-site modified the Ni-O and Ti/Zr-O boding and oxygen vacancies. The most interesting observation is the finding of indirect band gap around ~ 2.9 eV, this may be due to creation of defect levels and oxygen vacancies. Basically substitution on B-site by transition metal leads the cation octahedral ordering which alters the cation bonding, and therefore strongly affects the bandgap. Metastable states of $Ni^{+2}/N^{+3}$ ions create intrinsic oxygen vacancies/defects (as can be seen from the stoichiometry and matched with the EDAX and XRR data) which may develop the indirect bandgap in the system. The shift in $E_g$ may also be interpreted in context of the presence of impurities, [28] substitution ions concentration, [29] the level of structural and thermal disorder,[30] and native defects.[31] Most important, this indirect band gap helps to trap different wavelength of solar spectrum and hence may provide high $V_{oc}$ and $I_{sc}$. Choi et al demonstrated that the specific site substitution of ferroelectric bismuth titanate by Mott insulator lanthanum cobaltite leads reduction of band gap by at least 1 eV.[32] Thus, we have accomplished the small direct bandgap tunability ($\Delta E_g$ ~ 0.3-0.4 eV) and significantly large indirect band tunability ($\Delta E_g$ ~ 0.7-0.8 eV), which is a step forward for the classical ferroelectric PZT with transition metal-complex oxides.

Photoelectric effects depend on a number of factors such as, bandgap, intensity and frequency of the incident photons, absorption coefficient, carrier mobility and the domains (for bulk PV) interaction with light. We measured the photocurrent and transient current with one sun (power of solar simulator - 550 W) source in the metal-ferroelectric-metal (MFM) geometry. As shown in Fig. 5, the current-voltage behavior was investigated with a small bias field range; we found $V_{OC}$ in the range of (0.3-0.4 V) with short circuit current density ($I_{SC}$) in the range of 1-5 µA/cm$^2$ for PZTNi30/ITO/Glass and PZTNi30/LSMO/LAO respectively. The power conversion efficiency (PCE) is the key performance metric of an ideal solar cell, which is defined as follows: $\eta = \left(\dfrac{I_{max} \times V_{max}}{P_{in}}\right) = \left(\dfrac{FF \times I_{SC} \times V_{OC}}{P_{in}}\right)$ ……………………………………………………..(4)

where *Imax* and *Vmax* describe the bias current and voltage points where the photo-generated power reaches the maximum, $P_{in}$ is the power density of the incident light and *FF* is the fill factor. The PCE of the PZTNi30 based heterostructure is obtained from Fig. 5. It is about ~ 0.006 (+/- 0.004) % depending upon bias voltages and heterostructure configuration and its FF is 0.31 which is comparable to those obtained for perovskite oxides. [3,6] Polycrystalline samples showed exceptionally good switching of $V_{OC}$ under opposite polarity poling (±5 V) (see Fig. 5(b)); however, highly grain-oriented film had some in-built current even in dark without applying any voltage (see Fig. 5(a)) because it had some inbuilt polarization. The short circuit current density is much better for highly oriented films than polycrystalline films. These finding are comparable to ferroelectric BFO in MFM geometry, under similar growth and characterization conditions.[33]

The strength of photocurrents, persistency over a period of time, and transient behavior under ON and OFF illumination of light were examined in PZTNi30/LSMO/LAO(100) hetorostructure over different periods of time with 0 and ±10 V bias E-field (see Fig. 6). Sudden interruption and illumination of light allow the decay and growth of photo charge carriers over time. Growth and decay of photocurrent for ON and OFF states at different switching times were carried out under 0, and +/- 10 V bias E-field for short/long period of time (30/150 s) as can be seen in Fig. 6(a & b). Under these bias conditions, high photo-current density (0.1-0.5 mA/cm$^2$) and 1:4 to 1:5 ON and OFF current ratio were obtained with one second time period (experimental limit). An interesting feature in the transient currents can be seen in Fig. 6(c) and 6(d) during ON and OFF states, which exponentially increase or decrease with time, depending on the biasing conditions. This may be due to development of displacement current along or opposite to photo-charge carriers under bias E-field condition. These results indicate that the domain orientation and flipping with bias voltage are important factors for the bulk ferroelectrics photocurrent. These results are also suitable for opto-memory applications. Ferroelectric oxides have very slow charge carriers compare to the Si-based or organic photovoltaic devices. [34] Under bias E-field, charge carriers in polar oxides took long time to grow and decay with long saturation time. In this regards, present investigation illustrates sharp growth and decay of photo-charge carriers within the experimental limitations.

**Conclusions:**

In summary, we have successfully grown PZTNi30 single phase bulk photovoltaic ferroelectrics with switchable domains and photocurrents at nano/micro-scale. Substitutional modification by transition metal at Ti/Zr-site of PZT leads a decrease in direct and indirect band gaps without loss of its ferroelectric polarization. Experimentally, we showed that the cation modification of oxygen octahedra significantly reduces the indirect bandgap compared to direct bandgap. Photovoltaic effects are observed with significant amount of $V_{OC}$ (0.3-0.4 V) and good $I_{SC}$ (1-5 $\mu A/cm^2$); effect of poling and domains switching can be seen in the photocurrent and $V_{OC}$ performance. Thus, our investigations lead to the opportunities for more successful modification of ferroelectric materials for bulk photovoltaic effects and may be useful for opto-memory and energy applications.

**Acknowledgments**

This work was supported by the DOE grant DE-FG02-08ER46526, utilizing infrastructure support by NSF-RII-0701525.10

## List of Figures

Figure 1. Room temperature x-ray diffraction of PZTNi30 thin films grown on LSMO (α) coated LAO (*) substrate. Inset shows the PZTNi30 thin film grown on ITO/Glass (s) substrate.

Figure 2. (a) AFM topography images with an area of 5 x 5 μm$^2$ of PZTNi30/LSMO/LAO thin film structure (total thickness ~400 nm) before poling. PFM phase and amplitude image under different poling conditions; (b) phase and (c) amplitude images at ± 9 V poling of the respective areas.

Figure 3. Room temperature ferroelectric hysteresis loop for: (a) PZTNi30/LSMO/LAO and (b) PZTNi30/ITO/Glass thin film heterostructures. Insets show the respective dielectric constant (ε) versus bias electric field (E) curve for the respective sample.

Figure 4. (a) Transmittance, (b) Direct band gap and (c) Indirect band gap of PZTNi30/ITO/Glass structures, band gaps are calculated with transmittance data and Tauc's relations.

Figure 5. Current density versus voltage curves for (a) PZTNi30/LSMO/LAO for small voltage range, inset show large voltage range. (b) PZTNi30/ITO/glass thin film structures in dark and light without applying any voltage and under light illumination after negative and positive 5V poling (P) for PZTNi30/ITO/Glass.

Figure 6. Current density as a function of time of PZTNi30/LSMO/LAO structures with light (1 sun) ON and OFF state for the time periods (a) 30 s, and (b) 150 s without applying any voltage (c) 150 s , with applying +10 V (d) 150 s , with applying -10 V.

**List of Tables**

Table 1. Room temperature lattice parameters for bulk Pb(Zr$_{0.14}$Ti$_{0.56}$ Ni$_{0.30}$)O$_3$ (PZTNi30), polycrystalline and oriented PZTNi30 thin films deposited on ITO/glass and LSMO/LAO substrates respectively.

| PZTNi30 | Room temperature lattice parameters | |
| --- | --- | --- |
| | $a$ (Å) | c (Å) |
| Bulk | 3.9639 | 4.1279 |
| Polycrystalline Film | 3.9834 | 3.9939 |
| Oriented Film | 3.9899 | 4.0038 |

**References**

[1] V. Fridkin "Ferroelectric Semiconductors" Consultants' Bureau, New York 1980.

[2] James F. Scott, Carlos A. Paz de Araujo, Science, **246**, 1400 (1989).

[3] I. Grinberg, D. V. West, M. Torres, G. Gou, D. M. Stein, L. Wu, G. Chen, E. M. Gallo, A. R. Akbashev, P. K. Davies, J. E. Spanier & A. M. Rappe, Nature, **503**, 509, (2013).

[4] L. Pintilie, I. Vrejoiu, G. Le Rhun, and M. Alexe, J. Appl. Phys. **101**, 064109 (2007).

[5] G.Y. Gou, J.W. Bennett, H. Takenaka, A.M. Rappe, Phys. Rev. B. **83**, 205115 (2011).

[6] T. Choi, S. Lee, Y. Choi, V. Kiryukhin, and S.-W. Cheong, Science **324**, 63 (2009).

[7] Steve M. Young, Fan Zheng, and Andrew M. Rappe. Phys. Rev. Lett. **109**, 236601 (2012).

[8] M. Ichiki, Y. Morikawa, Y. Mabune, T. Nakada, K. Nonaka. R. Maeda. Microsyst Technol **12**, 143 (2005)

[9] S. Yang, J. Seidel, S. J. Byrnes, P. Schafer, C.-H. Yang, M. Rossel, P. Yu, Y.-H. Chu, J. F. Scott, J.W. Ager, L. Martin, and R. Ramesh. Nat. Nanotechnol. **5**, 143 (2010).

[10] V. K. Yarmarkin, B. M. Gol'tsman, M. M. Kazanin, and V. V. Lemanov. Phys. Solid State. **42**, 511 (2000).

[11] J. W. Bennett, I. Grinberg, and A. M. Rappe, J. Am. Chem. Soc. **130**, 17409 (2008).

[12] J. W. Bennett, I. Grinberg, P. K. Davies, and A. M. Rappe. Phys. Rev. B. **82**, 184106 (2010).

[13] K. K. Uprety, L. E. Ocola, and O. Auciello. J. Appl. Phys. **102**, 084107 (2007).

[14] Y. Inoue, K. Sato, H. Miyama. J. Phys. Chem. **90**, 2809 (1986).

[15] A. M. Glass, D. V. D. Linde, and T. J. Negran, Appl.Phys. Lett. **25**, 233 (1974).

[16] P. S. Brody, Solid State Commum. **12**, 673 (1973).

[17] J. Wang et al., Science **299**, 1719 (2003)

[18] S. R. Basu, L.W. Martin, Y.H. Chu, M. Gajek, R. Ramesh, R. C. Rai, X. Xu, and J.L. Musfeldt. Appl. Phys. lett. **92**, 091905 (2008).

[19] A. J. Hauser, J.Zhang, L. Mier, R. A. Ricciardo, P. M. Woodward, T.L.Gustafson, L. J. Brillson, and F. Y. Yang Appl. Phys. Lett. **92**, 222901 (2008).

[20] S. Y. Yang et al. , Appl. Phys. Lett. **95** 062909 (2009).

[21] D. Cao, J. Xu, Liang Fang, W. Dong, F. Zheng, and M. Shen. Appl. Phys. Lett. **96**, 192101 (2010).

[22] Y. Li, V. Nagarajan, S. Aggarwal, R. Ramesh, L. G. Salamanca-Riba, and L. J. Martínez Miranda. J. Appl. Phys. **92**, 6762 (2002)

[23] S.Gariglio, N.Stucki and J-m Triscone, Appl. Phys. Lett. **90**, 202905 (2007).

[24] G. Catalan, J. Seidel, R. Ramesh, and J. F. Scott, Rev. Mod. Phys. **84**, 119, (2012).

[25] J. Tauc. Mat. Res. Bull. **3**, 37 (1968).

[26] R. Ardebili, J. P. Charles, L. Martin, J. Marucchi, and J. C. Manifacier, Mat. Res. Bull. **25**, 1407 (1990).

[27] Bin Chen, Zhenghu Zuo, Yiwei Liu, Qing-Feng Zhan, Yali Xie, Huali Yang, Guohong Dai, Zhixiang Li, Gaojie Xu, and Run-Wei Li, Appl. Phys. Lett. 100, 173903 (2012)

[28] M. Tanenbaum and H. B. Briggs. Phys. Rev. **91**, 1561 (1953).

[29] R. Pino, Y. Ko, P. S. Dutta, S. Guha, L. P. Gonzalez. J. Appl. Phys. **96**, 5349 (2004).

[30] G. D. Cody, T. Tiedje, B. Abeles, B. Brooks, and Y. Goldstein. Phys. Rev. Lett. **47**, 1480 (1981).

[31] E. Burstein, Phys. Rev. **93**, 632 (1954); T. S. Moss, Proc. Phys. Soc. (London), Sect. B **67**, 775 (1954).

[32] Woo Seok Choi, Matthew F. Chisholm, David J. Singh, Taekjib Choi, Gerald E. Jellison Jr. & Ho Nyung Lee, Nature communications | 3:689 | DOI: 10.1038/ncomms1690 (2012).

[33] R. K. Katiyar, A. Kumar, G. Morell, J. F. Scott, and R. S. Katiyar, Appl. Phys. Lett. **99**, 092906 (2011).

[34] Antonio Luque, and Steven Hegedus, Handbook of Photovoltaic Science and Engineering, John Wiely & Sons Ltd. (2003).

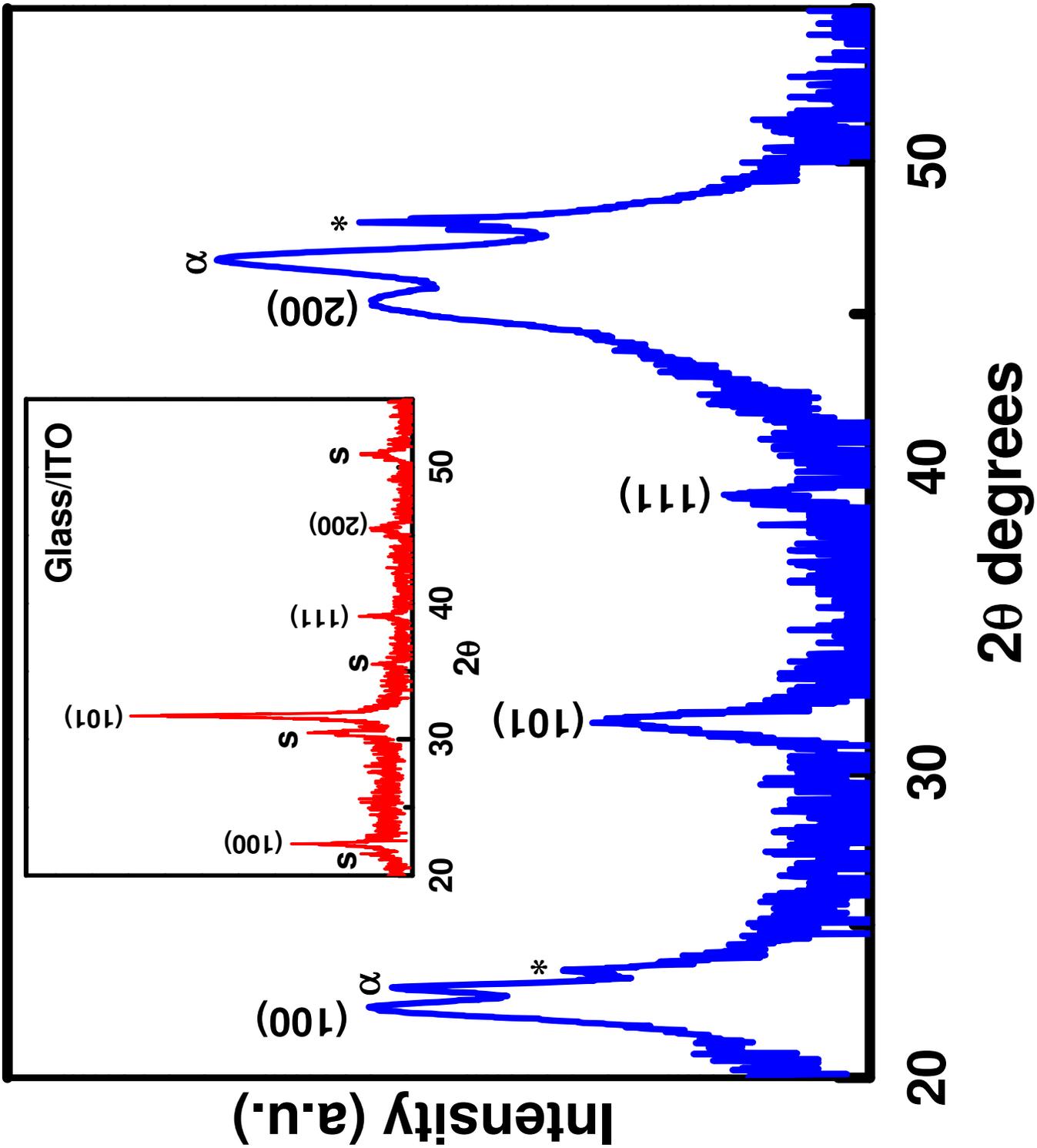

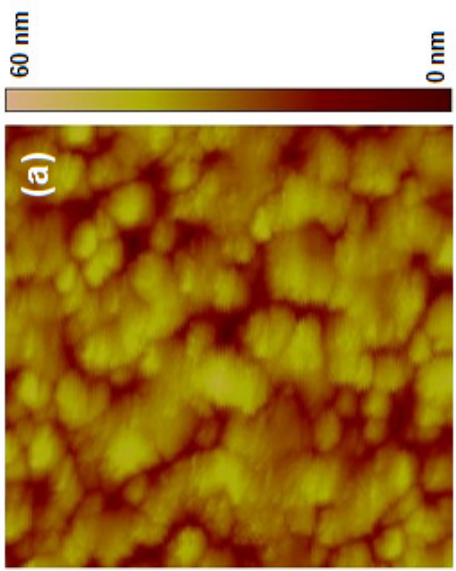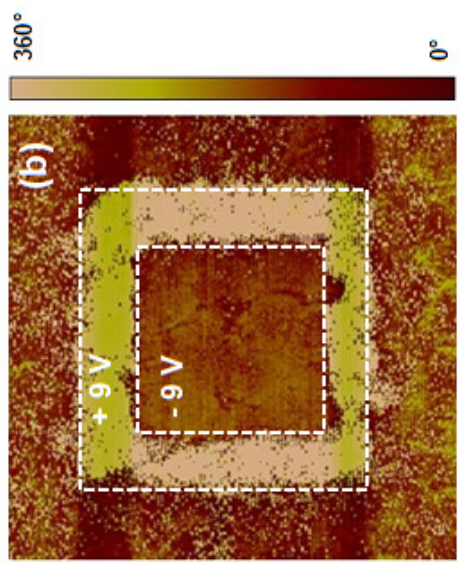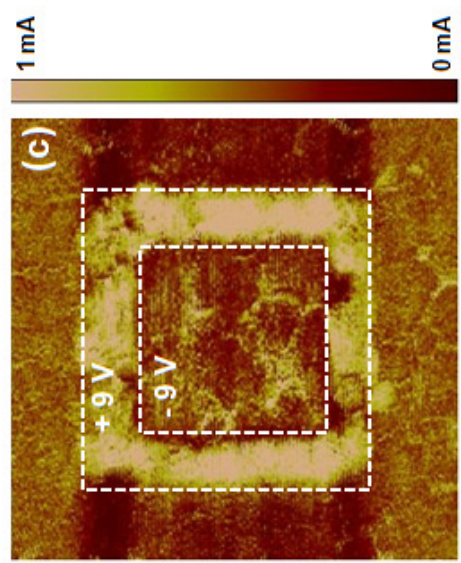

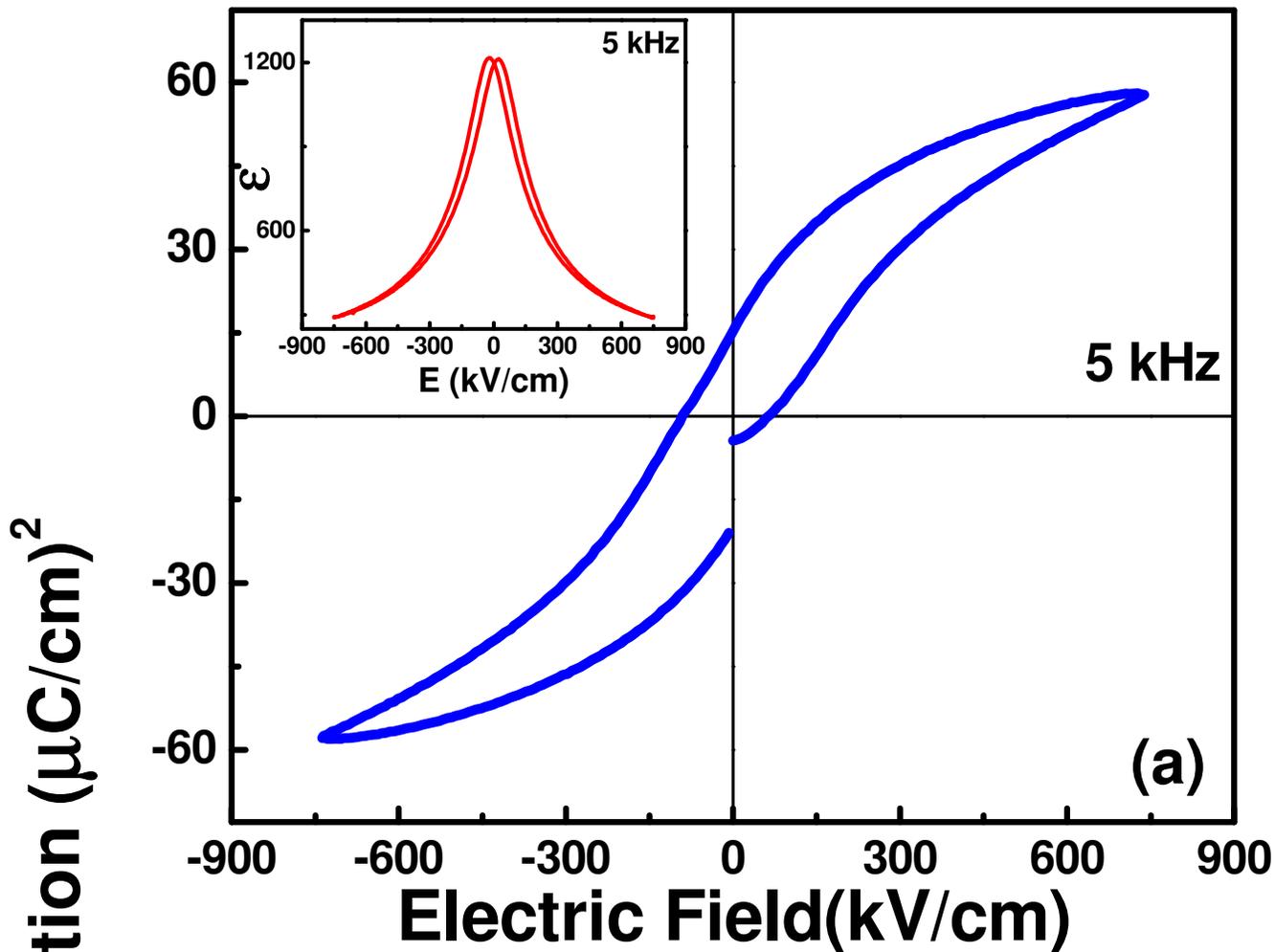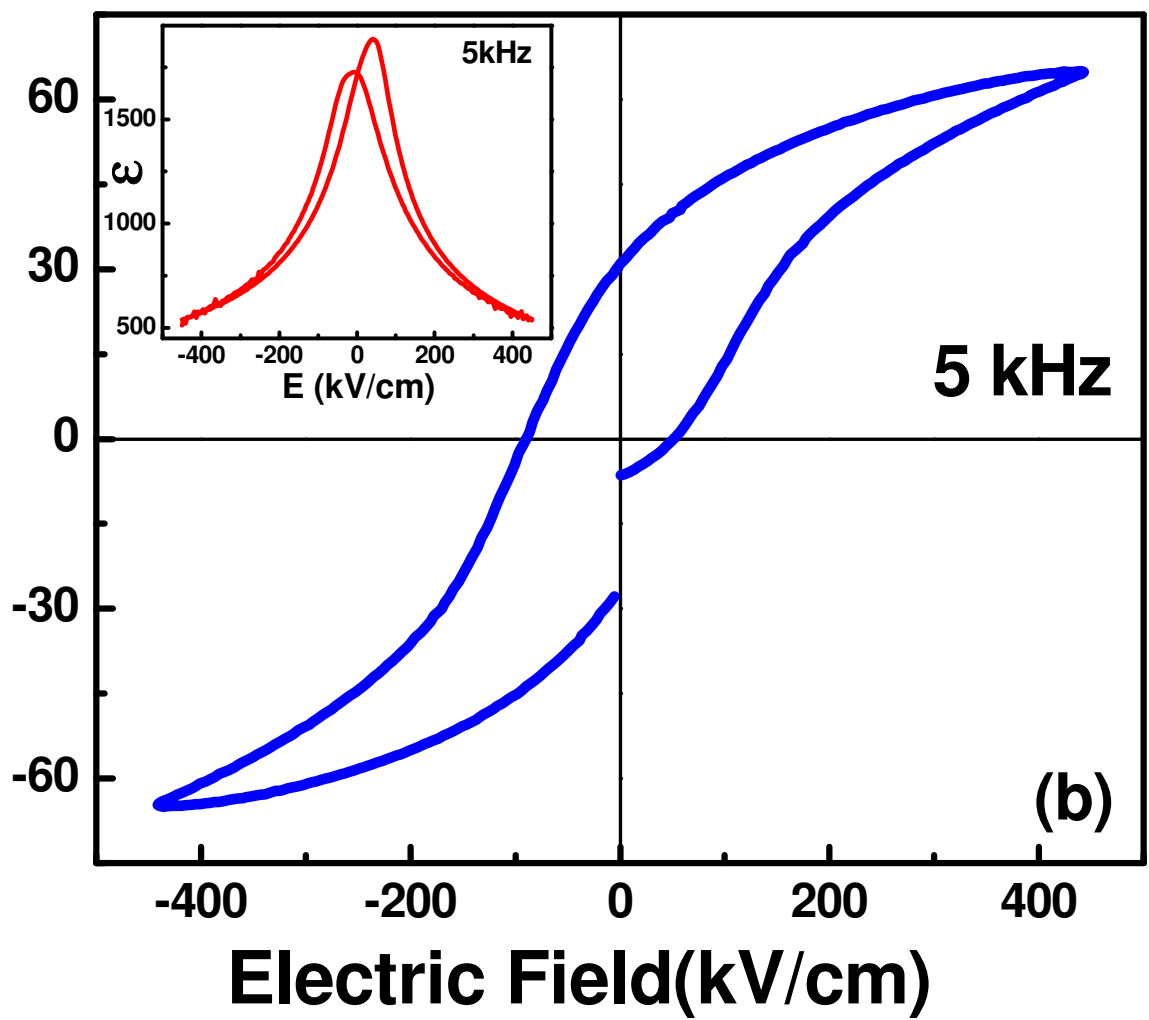

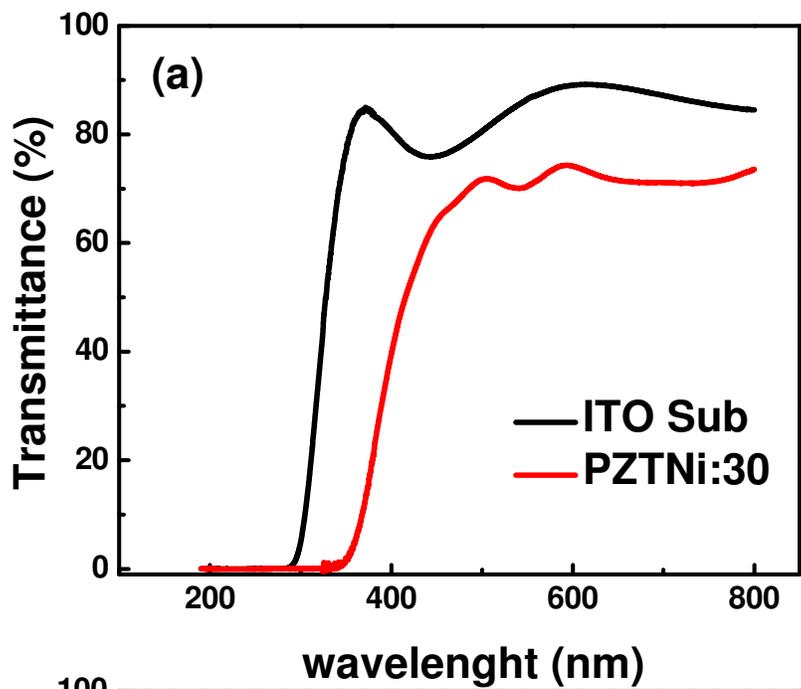

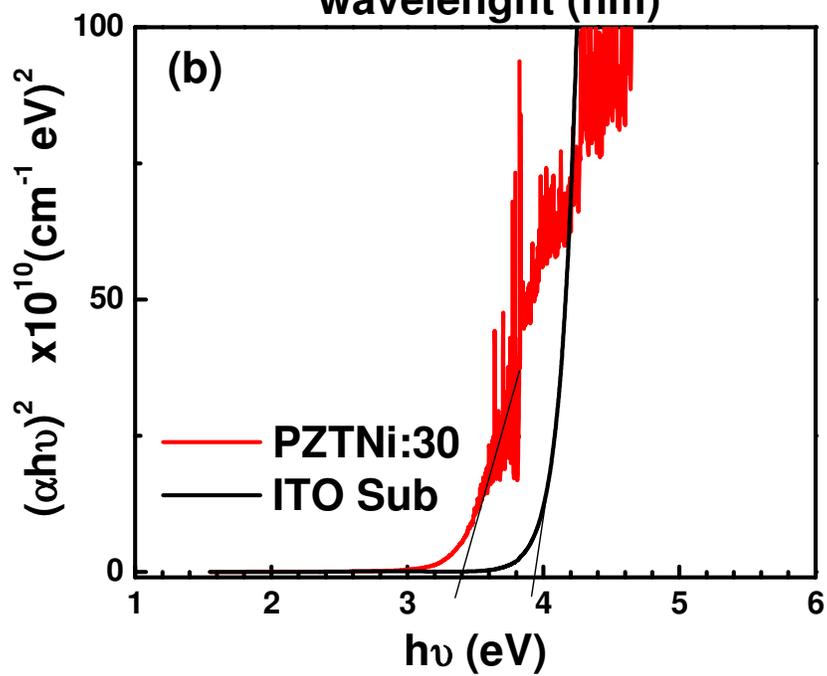

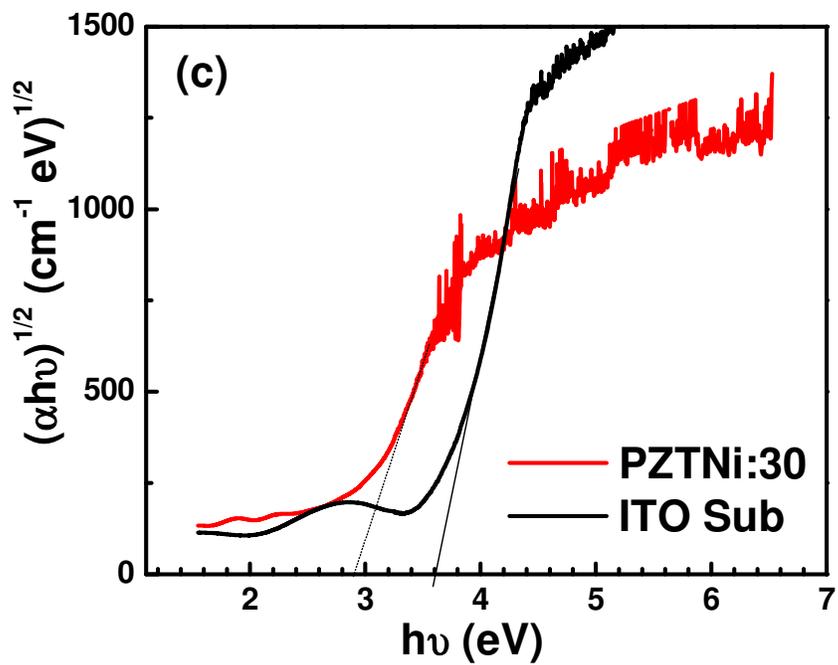

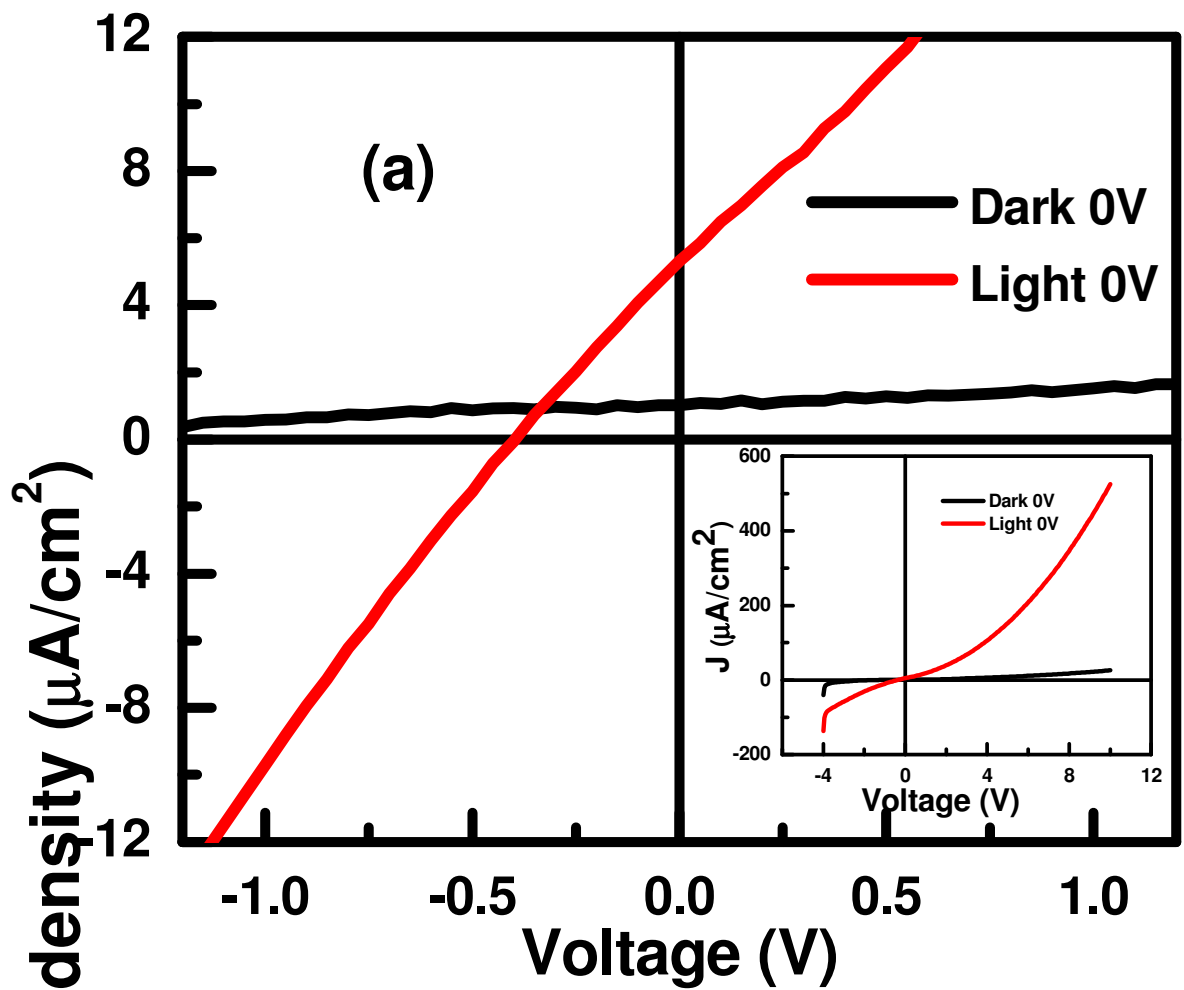
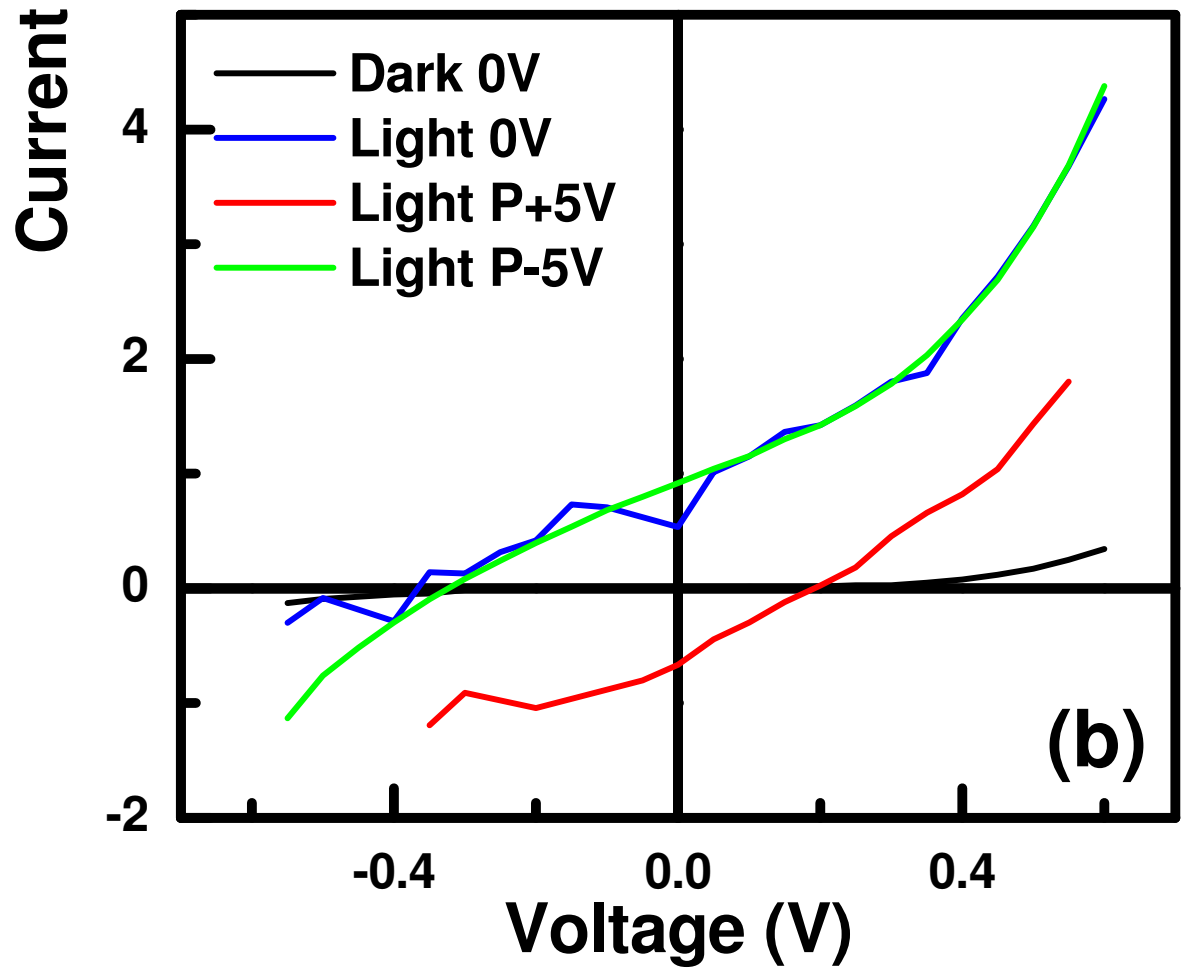

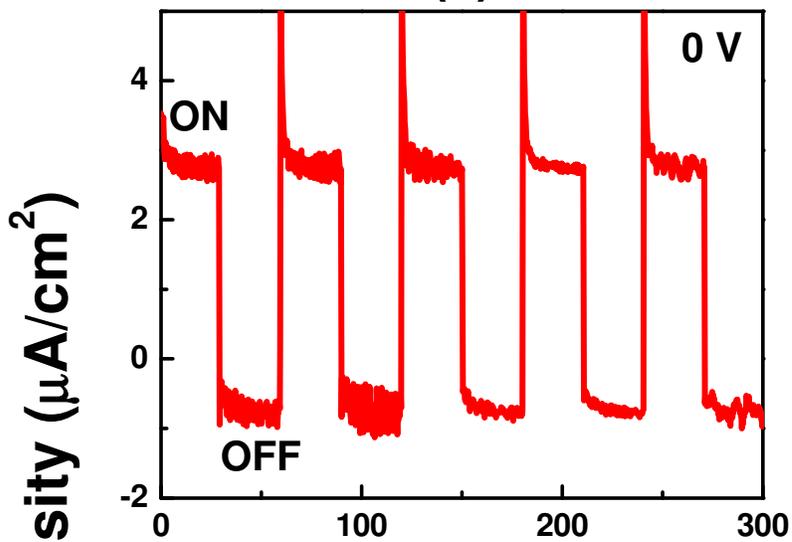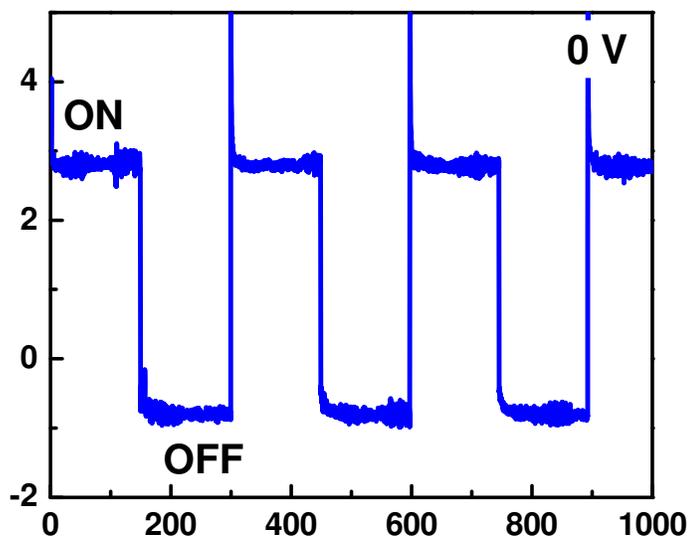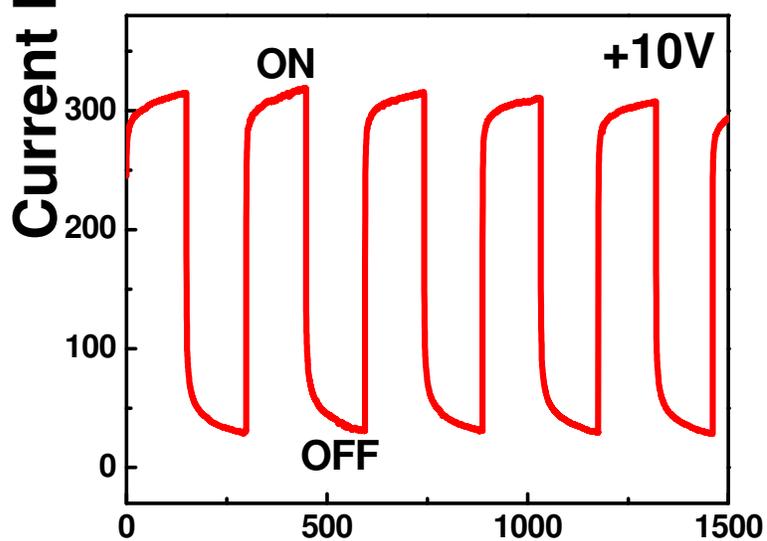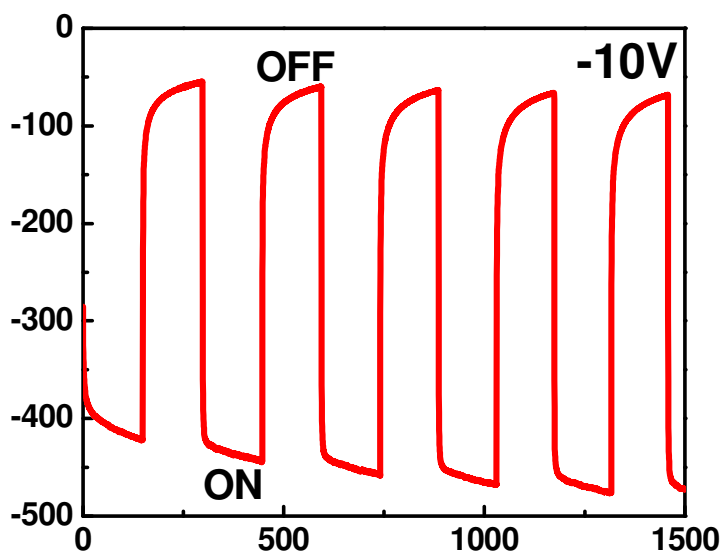